\documentclass[useAMS,usenatbib,usegraphicx]{mn2e}

\title{Variability and multi-periodic oscillations in the X-ray light
curve of the classical nova V4743 Sgr}
\author[E. Leibowitz, M. Orio, R. Gonzalez-Riestra,
Y. Lipkin, J-U. Ness, S. Starrfield, M. Still, E. Tepedelenlioglu]
{E. Leibowitz$^1$, M. Orio$^{2,3}$, R. Gonzalez-Riestra$^4$,
Y. Lipkin$^1$, J-U. Ness$^{5,6}$,\cr
 S. Starrfield$^6$, M. Still$^7$, E. Tepedelenlioglu$^8$\\
$^1$School of Physics and Astronomy, Sackler Faculty of Exact Sciences,\\
 Tel Aviv University, Israel\\
$^2$INAF -- Turin Astronomical Observatory, Strada Osservatorio 20,\\
 I-10025 Pino Torinese (TO), Italy\\
$^3$Department of Astronomy, 475 N. Charter Str., Madison WI
              53706, USA\\
$^4$XMM-Newton Scientific Operation Centre, ESAC,
 P.O.  Box 50727,\\ 28080 Madrid, Spain\\
$^5$ Department of Physics, Rudolf Peierls Centre for Theoretical Physics,\\
 University of Oxford, 1 Keble Road, Oxford OX1 3NP, UK \\
$^6$ Department of Physics and Astronomy, Arizona State University,\\
   Tempe, AZ 85287, USA\\
$^6$ South African Large Telescope, PO Box 9, Observatory 7935, South Africa \\  
$^8$ Physics Department, 1150 University Avenue, Madison WI 53706, USA}
\date{}

\pagerange{\pageref{firstpage}--\pageref{lastpage}}
\pubyear{}

\begin{document}

\maketitle

\label{firstpage}

\begin{abstract}
The classical nova V4743 Sgr was observed with XMM-Newton for
about 10 hours on April 4 2003, 6.5 months after optical maximum. At
this time, this nova had become the brightest supersoft X-ray source
ever observed. In this paper we present the results of a time series
analysis performed on the X-ray light curve obtained in this
observation, and in a previous shorter observation done with
Chandra 16 days earlier. Intense variability, with amplitude as
large as 40\% of the total flux, was observed both times.
Similarities can be found between the two observations in the
structure of the variations. Most of the variability is well
represented as a combination of oscillations at a set of discrete
frequencies lower than 1.7 mHz.
At least five frequencies are constant 
over the 16 day time interval between the two
observations. We suggest that a periods in the power spectrum of
both light curves at the frequency of 0.75 mHz and its first harmonic
are related to the spin period of the white dwarf in the system, and
that other observed frequencies are signatures of nonradial white dwarf
pulsations. A possible signal with a  24000 sec period is also found in
the XMM-Newton light curve: a cycle and a half are clearly identified.
This period is consistent with the 24278 s periodicity discovered in the
optical light curve of the source and thought to be the orbital period
of the nova binary stellar system.
\end{abstract}


\section{Introduction}

Time series of astronomical observations are crucial for the
study of variable stars in general, and of Cataclysmic Variables
in particular.
For Classical Novae (CN), X-ray time series give us information
about temporal characteristics of very localized areas within
these stellar systems that
are inaccessible at other wavelengths. In particular, in a
classical nova system, in the X-ray range we may be observing
directly the surface of the white dwarf (WD) or the very inner region
around it. This gives us a direct
look into the temporal behaviors of the site where the
thermonuclear outburst takes place.

In this paper we report results obtained from our analysis
of X-ray light curves (LC), using data obtained with the XMM-
Newton telescope for the classical nova V4743 Sgr (N Sagittarii
2002 No. 2), about 6.5 months after the nova outburst.
We also reanalyze the X-ray light curve of the nova observed
with Chandra two weeks earlier (see Ness et al 2003).
The results we present here provide us with details of information
that is hardly obtained by any other means, underlining the great
potential of  time series applied to X-ray observations.
\section{Nova V4743 Sgr}

Nova V4743 Sgr (Nova Sgr 2002 no. 2) was discovered in outburst in September of
2002 and reached V=5 on 2002 September 20  (Haseda 2002). It was a very
fast nova, with a steep decline in the optical light curve and large ejection
velocities. The time to decay by 3 mag in the visual (t$_3$),  was 15 days and the
Full Width at Half Maximum (FWHM) of H$\alpha$ line reached 2400 km s$^{-1}$
(Kato 2002). An estimate of the distance based on infrared observations is
 $\approx$6.3 Kpc (Lyke et al. 2002).

In December 2002, the nova was observed for the
first time with Chandra ACIS-S.
At that time the nova was a very soft and moderately luminous X-ray
source, with a count rate about 0.3 cts s$^{-1}$. There were indications that
the  nova was not at the peak of X-ray luminosity yet,
so a  Chandra-LETG grating observation was done on
2003 March 20. A description of the instrument is found in Brinkman et al.,
(2000). The count rate was astonishingly high, 40 cts s$^{-1}$  during 3.6 hours,
then a slow decay, lasting for an hour and a half, was followed by
another 1.5 hours of very low luminosity with  a measurement of only
$\approx$0.02 cts s$^{-1}$ (Ness et al. 2003, Starrfield et al. 2003).

A decline of a supersoft X-ray source in a nova was observed once before
with BeppoSAX, in V382 Vel (Orio et al. 2002). The reason for the sudden
decline remains unexplained. Before the decline, the unabsorbed
flux in the Chandra observation
of V4743 Sgr was close to 10$^{-9}$ erg cm$^{-2}$ s$^{-1}$ and the spectrum
 was extremely soft,
with atmospheric absorption features (Rauch et al. 2005, Petz et al.
2005). During the high luminosity and the decay phases, a periodicity of
1324 s was detected, with fluctuations of 20\% of the mean count rate
(see Starrfield et al. 2003, Ness et al. 2003).

Wagner et al. (2003) discovered a period of 24278$\pm$259 s
in the optical light curve of the star. They proposed that
this is the orbital period of the nova binary stellar system.
The $\simeq$24000 sec periodicity was measured again in the optical LC of
the star in July 2004 (Wagner et al, 2006). These authors found
also a period of 1341 sec in that LC as well as a period of 1419
sec which seems to be a beat period of the above two.

\section{A new observation with XMM-Newton}

A new observation program was proposed by us to the  XMM-Newton
Project Scientist as a target during the Discretionary Time. The nova was
observed with this telescope on April 4 2003, for 10 hours (see Orio et al.
2003).
A description of the mission can be found in Jansen et al. (2001).
Three X-ray telescopes with five X-ray detectors were all used:
the European Photon Imaging Camera pn (see Str\"uder et al.
2001), two  EPIC MOS cameras (Turner et al. 2001),  and two Reflection
Grating Spectrometers (RGS-1 and  RGS-2, (see den Herder et al.
2001). The observation lasted a
little over 36000  seconds with EPIC and the RGS-1 and RGS-2
overlap for 35306 seconds.
The satellite carries also a UV/Optical telescope (OM; Watson et al.
2001). Observations with this instrument where carried out in
imaging mode with the UVW1 filter (effective wavelength: 291 nm).
35 exposures were done
 with exposure times 800 s, which were too long for the timing analysis done in this
 article. The average count rate was 252$\pm$10 cts s$^{-1}$, and
 we measured an average flux 1.2 $\times 10^{-13}$ erg cm$^{-2}$ s$^{-1}$.
The RGS spectra are discussed in detal in a forthcoming article (Orio
 et al. 2006, in preparation).
 
The data obtained by the two  EPIC MOS cameras suffered from very severe pile up
effect that made them unsuitable for our analysis. The data
initially collected by
the EPIC  pn camera, operated in the prime mode, with the
full window, and the thin filter, suffer from the same effect
during the first 79 minutes of observation. The operation of
the pn camera was then switched to timing mode to avoid
 severe pile-up. Here we use only the data obtained in timing mode,
 disregarding the
first $\approx$5000 seconds of the EPIC-pn imaging mode observations.
The measured
average, background corrected count rate in timing mode
was 1309.5$\pm$0.3 cts s$^{-1}$ (Orio et al. 2003). The variations  are by
up to 40\%.

The data were reduced with the  ESA XMM Science Analysis System (SAS)
software, version 5.3.3., using the latest calibration files available
in July of 2005.  We further subdivided
the  EPIC-pn data into three different ``supersoft''
distinct energy bands: low: 0.2-0.4 keV, medium: 0.4-0.6 keV and high: $>$0.6 keV
The RGS count rates are about 57 cts s$^{-1}$.  The unabsorbed flux was
measured by us to be  1.5 $\times$ 10$^{-9}$  erg cm$^{-2}$ s$^{-1}$,
 consistent with the flux measured 16 days earlier with Chandra.

\section{The two X-ray Light Curves}

\subsection{The April 2003 XMM light curve}

The XMM data reduced from the various instrument on board the
satellite
as described above, provided us with six X-ray LC of the source,
of which five
are independent of each other. These are the 2 RGS and the 3
``color''
EPIC-pn LC.  The 6th LC is the integrated
EPIC-pn LC. The optical/UV data was not useful for the analysis
presented
in this paper since the exposures were 800 s long.

The data extracted from the  RGS-1 and RGS-2 grating were binned
into
0.574 s wide bins. This is the time it takes to read each one of
the eight
CCDs that together collect the photons of the whole spectrum. The
RGS
dispersion gratings cover the  5-35 \AA \ wavelength range (0.35-
2.5 keV),
although we obtained useful signal only in the 26-35 \AA \ range.
The RGS data are also piled-up, but in a dispersion instrument,
piled-up
events at a discrete wavelength increase in pulse height amplitude
by an integer
multiple of the intrinsic energy. Furthermore, since the softness
of the source
precludes any intrinsic photons from higher spectral orders, we
can confidently
identify events that occur within the higher order spectral masks
for an on-axis
point source as piled-up first order photons.  This is verified by
line matching
of the piled-up events using the first order response matrix.
Source events
dominate over background and scattered source light in the first
three orders.
Consequently we added  events within the second and third order
extraction masks
to the first order events, thus reclaiming the piled-up events and
increasing
the signal-to-noise of the spectra.

The LC of the  EPIC-pn camera in timing mode cover the last 30000
seconds of the two  RGS light curves. The time resolution of  EPIC-
pn
in this mode is 0.03 msec, but the light curve turned out to have
many short (few seconds)
time intervals excluded from the {\sl Good Time Intervals (GTI)},
mainly because
the source count rate was too high for the available telemetry
band width, and
as a result part of the data were lost. The measured  pn light
curve is
thus split into many portions lasting  a few seconds each, with
``holes'' of few
seconds duration in the light curve. The energy range of the  pn
camera is
0.2-10 keV, and the light curve we present and discuss in this
paper was
extracted with 1 second wide time bins. The broad band  pn data
were
subdivided into the three  narrow band light curves described
above : LOW, MED,  and HIGH.
\begin{figure}
\includegraphics[width=87mm]{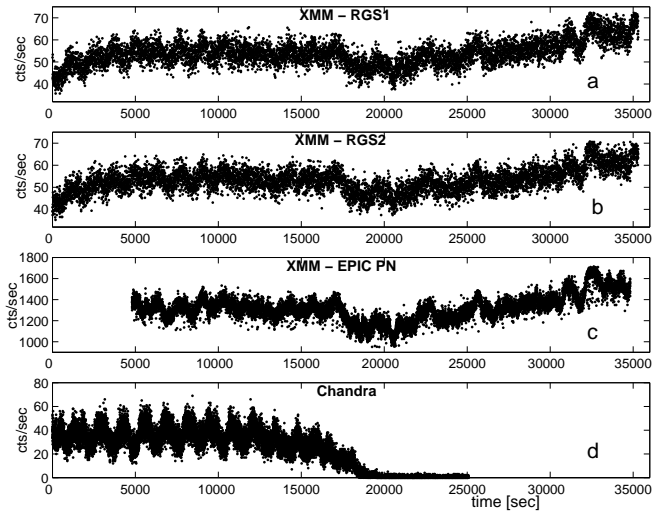}
\caption{  X-ray light curves of V4743Sgr: (a) XMM-Newton 
RGS-1 grating, b) XMM-Newton RGS-2 grating, (c) XMMNewton/
EPIC-pn, d) Chandra LETG grating. The Chandra observations 
were performed on March 19 and those of XMMNewton 
on April 4 2003.  }
\label{fig1}
\end{figure}
Fig. 1a, 1b and 1c display the two RGS light curves and the broad
band
pn one. The three narrow band  pn LC have the same structure as
that of
the broad band one.
All six light curves show similar variations. The
similarity between them is further manifested in the
structure of their power spectra (PS). Within the statistical
uncertainty, all the peaks in the PS that rise  higher than  1
$\sigma$ of the noise appear at the same frequencies in all the
six power spectra (but see Section 6.4). Exceptions are peaks at
very low frequencies, corresponding to periodicities of the order
of the observing time duration itself, i.e. P$\simeq$36000 sec.
Those for the RGS LC are obviously different from those in the pn
LC.

In order to minimize possible instrumental systematic errors or biases
in the data to be analyzed, we constructed a general XMM-Newton light
curve by taking an average LC of the two different instruments on board
the satellite. We computed the mean LC of the two RGS cameras, and then
the mean of that LC and the integrated epic-pn LC. The units of the
latter were normalized to fit those of the former in the 30000 sec
overlap time between the two. The upper, thick curve in Fig. 2a presents
this LC, binned into 600 equally spaced bins, each one 58.8415 sec wide.
A linear trend was removed from the data by least squares procedure.

The y (count-rate) value of each point in the binned LC is the average y
value of all points within the corresponding bin. The formal statistical
error in that average value is the StD of all bin points divided by the
square root of the number of points in the bin. For the first 6000 sec
of the LC, for which we have only the RGS cameras data, this procedure
gives an estimated error of 0.63 cts s $^{-1}$. For the rest of the LC seen in
the upper curve of Fig. 2a, the estimated error in each point is 0.46
cts s $^{-1}$.

\subsection{The March 2003 Chandra LC}

V4743 Sgr was observed by the Chandra X-ray telescope 16.5 days
prior to the XMM-Newton observations. Details and analysis of this
run are given by
Ness et al (2003). Fig. 1d is the Chandra  LC in bins of 1 s.

The decline effect discussed in the above references is clearly
evident. Fig. 2b shows the LC in the first
18000 s of the Chandra run. A polynomial of third degree fitted to
the data
by least squares is subtracted from the data to remove the long
range trend
imposed on the LC by the decline. The data are binned into 300
bins,
each one is 60 s wide. We estimate that the error in each point in
Fig. 2b is
0.8 cts s$^{-1}$.

\section{Time series analysis: the power spectrum}

Following the formalism outlined by Scargle (1982) we computed
the power spectra (PS) of the 2 LC shown
in Fig. 2, and present them in Fig. 3a and 3b. The PS cover
the period range from 36000 s, the duration of the XMM Newton
observations, down to 300 s, corresponding to about
2/5 of the Nyquist frequency of both LC. The computation
was performed on a grid of 3000 equally spaced frequencies
which over-samples the PS frequency range by a factor 25.
Both PS appear to consist of a set of peaks at discrete
frequencies
with no apparent continuum. We also checked the
frequency space up to the Nyquist frequency itself and found
a flat, near zero PS there.

We divide the frequency range of the two PS into 3
regions: a Red (R) region of frequencies smaller than 0.167
mHz (period 6000 s), a Medium (M) range of 0.167-1.7 mHz
and a Blue (B) range of frequencies higher than 1.7 mHz
(period 590 s). These ranges are indicated on Fig. 3 and we
shall refer to them in the following sections.

\begin{figure}
\includegraphics[width=87mm]{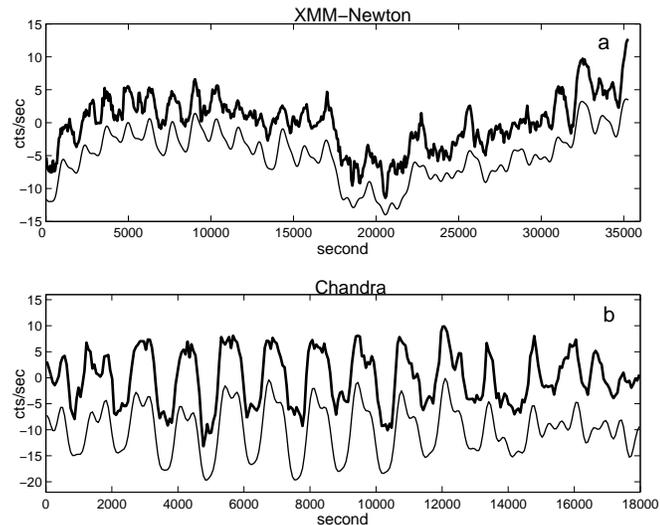}
\caption{ a) The thick curve is the X-ray light curve of V4743 
Sgr obtained by binning the mean measurements of the RGS1, RGS2 and EPIC-PC detectors on
board the XMM-Newton satellite into 600 equally spaced bins. The thin curve was created by 
Fourier series of 12 frequencies best fitted to the data by least 
squares. This curve is lowered by 5 units to allow comparison 
with the observed one. b) The thick curve is the Chandra light 
curve in the first 18000 s, binned into 300 bins. The thin curve 
shows a 6 term Fourier representation of the observed light curve
(lowered by 10 units).}
\end{figure}
\begin{table}
 \centering
 \begin{minipage}{85mm}

  \caption{List of the 12 most significant periods found in the LC
of V4743 Sgr as measured by XMM-Newton, along with their
respective amplitudes. The
numbers in the 4th column are labels assigned to the corresponding
periodicities
discussed in the text. b) The same for the six most significant
periodicities in the Chandra LC.
The LC obtained with the XMM-Newton 12 term series and the 6 term
Chandra series are shown as the thin curves in figures 2a and 2b.}

  \begin{tabular}{@{}lrrrr l@{}}
Frequency &   Period     & Amplitude &  Number\\
  & & & \\
 0.039412 & 25373  &     4.4757  &  1 \\
  0.10143 & 9859.2 &     1.3735  & \\
 0.7633 & 1310.1 &    1.0578   & 3 \\
 0.1687 &  5927.6 &    1.0138 & 7 \\
 0.38063 & 2627.2    &  1.0134 & \\
 0.33316 & 3001.6    &  0.95226 & \\
 0.13138    & 7611.6    &  0.9179 & \\
 0.72908 & 1371.6   &  0.90292 & 2 \\
 0.30915  & 3234.7   &  0.73275 & 6 \\
 0.60606  & 1650      &  0.60696 & \\
 0.43798  & 2283.2    &  0.60621 & \\
  1.5015 & 665.98   &  0.51516 & 5 \\
   & & & \\
Table 1.b: Chandra  LETG: & & & \\
 & & & \\
0.7773 & 1286.5  &     4.7977  & 3 \\
 0.72108 & 1386.8   &    3.7524  & 2 \\
 1.4817  & 674.92   &   1.7495  & 5 \\
 2.2425  & 445.94     &  1.4742  & \\
   0.3139   & 3185.7    &   1.4129  & 6 \\
  0.20202  & 4950.0    &   0.9792  & 7 \\
  & & \\
\end{tabular}
\end{minipage}
\end{table}

\begin{figure}
\includegraphics[width=87mm]{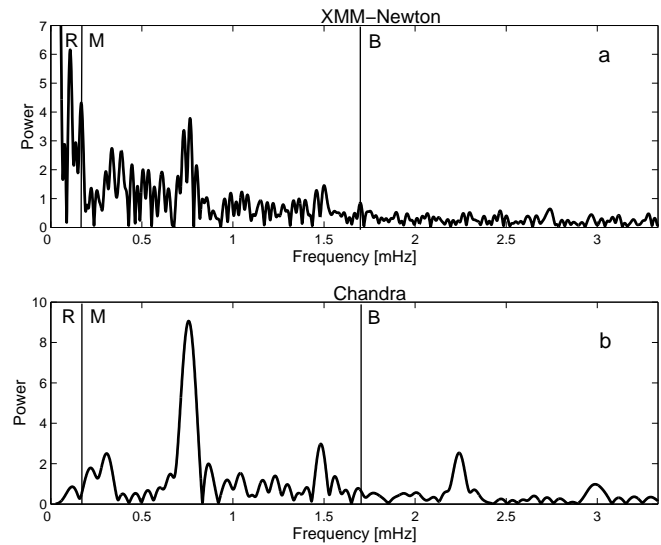}
\caption{Upper panel: power spectrum of the XMM-Newton 
light curve shown in Fig. 2. Lower panel: power spectrum of the 
Chandra LC shown in Fig. 2. The frequency regions R, M and B 
are discussed in the text.
}
\end{figure}
\subsection{The R Spectrum}

The clear  difference between the two PS in the extreme red
end of their frequency range reflects the obvious difference in
the broad structure of the LC shown in Fig. 2. This difference is
mostly due to the decline of the emission during the Chandra
run, mentioned in Section 2, and the resulting detrending
process that we had to apply to the Chandra LC. Due to the
decline, the Chandra data do not provide us with
information on the variability of the source in March 2003,
on time scale longer than 6000 s, other than the 
decline itself.

The 36000 s long XMM-Newton observations of April,
however, do show significant variability on a longer time
scale. This is evident by inspection of Fig. 1 and 2, and
by the high peaks in the R end of the XMM-Newton PS
shown in Fig. 3. The highest peak in this band and in the
entire PS is around a frequency f1=0.041 mHz. The corresponding
period is P1=24300 s. Within the uncertainty in this number it is
consistent with
the 24380 s periodicity discovered by Wagner et al (2003) in
the optical light of the nova, as mentioned in Section 2.

\subsection{The M and B spectral ranges}

Except for the difference in the R region, the two PS shown
in Fig. 3 reveal a distinct qualitative similarity between
them. The peaks in the M region are all higher than the
peaks in the B band. The only two exceptions are the 3rd
and the 4th harmonics of the prominent 0.75 mHz feature
seen in both LC and discussed in the following section. In
fact, none of the frequencies of the 20 highest peaks in the
XMM-Newton PS is in the B band. In the Chandra case,
among the frequencies of the 18 highest peaks, only the two
harmonics of the 0.75 mHz features are in the B band and all the
rest are in the M band. Using the binomial statistics formula we
find that for a random uniform distribution of frequencies in our
M+B band,
the probability of the apparent distribution in the XMM-Newton PS is
 1.8 $\times 10^{-5}$ and for the Chandra PS it is 6.5 $\times 10^{-4}$. 
The similarity between these two observed highly improbable
frequency distributions, in the light curves obtained using entirely
independent measuring devices and reduction procedures, indicates
that the origin of the apparent
variations in the two LC is in the source itself. Furthermore,
this similarity proves that it is very likely that oscillations in X-ray
flux maintained their characteristic time scale of a few thousand
seconds for at least 16 days. We shall now show that
precise matches of specific frequencies can also be found between
the two LC, and we shall make a quantitative estimate
of their statistical significance.

\section{Frequency matching}

\subsection {Oscillation frequencies}

In the M and B regions, both PS are dominated by a high
feature around the frequency of 0.75 mHz. It is statistically
highly significant in both PS, as evident by its large height
over the noise level. Quantitatively, one can show by 
bootstrap analysis (Efron \& Tibshirabi 1993) that in both
cases, the probability for obtaining such a high peak as a
result of a random process is negligibly small ($< 10^{-5}$). The
second harmonic of this feature around the frequency 1.5
mHz is also clearly seen in the two PS. The 3d and 4th
harmonics around 2.25 and 3 mHz are also prominent in
the Chandra PS, where they are also statistically significant
features. The appearance and the prominence of the 0.75
mHz feature in the two PS indicate that one, or possible two
cycles around this frequency probably persisted in the X-ray
LC of the nova for at least 16 days, the duration of time
between the Chandra and the XMM-Newton observations.

Examining only their height in the PS, all the peaks
other than the 0.75 mHz and its harmonics turn out not to be 
statistically very significant. Each of the two observations lasted for
no longer than a few to a few tens characteristic times of the
apparent oscillations in the M frequency
band. This is why examining the  PS we cannot decisively distinguish
between random noise on a time scale corresponding to this
frequency band and truly periodic or quasi-periodic variability of
the light source itself. However, the two X-ray observations were
16 days apart. If we find oscillations of considerable amplitude
that have the same frequency in both LC, their statistical
significance can be evaluated by calculating
the probability of random occurrence of such coincidences.

The frequencies of peaks in the PS of a LC, especially when it is
rich with many densely populated ones, may have a systematic
error. In the computation of a PS of a time series,
the power in each frequency measures the fitness to the data of a
single harmonic oscillation with the corresponding frequency,
regardless of the behavior of the LC at any other
frequency. In a finite noisy time series, if the variation is
also modulated by another, nearby frequency, the two frequencies
may interfere.   Any attempt to compare the characteristic
frequencies of our 2 LC
is sensitive to this possible error. In order to improve the
accuracy of the frequency determination we therefore applied a
least squares fitting  process to the data. We used the frequency
values of the highest peaks in the PS as our initial step, and by
varying their values we found the set of frequencies that produces
a Fourier series that fits the observed LC best in the least
square sense.

Table 1 presents the 12 frequencies and periods of the highest
amplitudes in a Fourier presentation of 12 or more terms that fits
best the XMM-Newton light curve. It also lists the 6 frequencies
of highest amplitude in the Fourier presentation of 6 or more
terms of the Chandra LC. The table lists the values in a
descending order of the corresponding amplitudes. The fourth
column
presents the labels by which we designate the periods in this
paper. We note that all the 12 XMM-Newton frequencies listed in
the table also coincide, as explained in detail in the next
section, with frequencies that are among the 17 highest peaks in
the PS of that LC. In the Chandra case, except for the B frequency
of the 445 sec period, all other 5 M frequencies are among the
frequencies of the highest 10 peaks in the PS of that LC.

The lower, thin curves in Fig. 2a and 2b are plots of the
Fourier series of the 12 and the 6 frequencies listed in Table 1.

\subsection{Coincidences}

Looking for coincidences between frequencies of the XMM-
Newton LC with frequencies in the Chandra LC, we must first define
what we consider a coincidence. An uncertainty in the value of the
frequency of an apparent oscillation in a noisy light curve is
1/T,
where T is the duration of the observing run. For a LC
oscillating with a given frequency, this is the half width at
zero level of the peak in the PS around this frequency. We
now define a frequency f(XMM) in the XMM-Newton LC
as coincident  with a frequency f(Chan) in the Chandra
LC if the distance between the two on the frequency axis is
smaller than the sum of the uncertainties in each LC divided
by K, namely, if

 $$ |f(XMM)-f(Chan)| < {{1 \over T(XMM)}+ {1 \over T(Chan)} \over
{\rm K}}=
 {\Delta f \over {\rm K}}$$
where T(XMM)=36000 s and T(Chan)=18000 s are the length of the
corresponding LC.

Comparing numbers in the table we find that all 5
Chandra M frequencies coincide (with K=2) with 5 frequencies among the 9
XMM-Newton M frequencies. The 5 coinciding periods are labeled 2-7 in
the table.
 Out of these 5, 4 are coincident with
K=4, namely, the distance between the
XMM-Newton and the corresponding Chandra frequency is
smaller than 1/4th of the sum of the uncertainties in the
frequency values.

\subsection{Probabilities}

The significance of the matches that we find between frequencies
of the XMM-Newton and the Chandra LC can be evaluated by
statistically testing the hypothesis that the two sets are
independent of each other. Under this assumption we can calculate
the probability of obtaining the number of observed matches. From
a uniform distribution of
numbers in the M+B band we select randomly a set of 6 numbers
for Chandra and a set of 9 numbers for XMM-Newton. We then find
the number of matches between these two sets, in the above
mentioned sense. By repeating the sampling many thousands times,
we find that the probability of obtaining 5 matches of the K=2
type, between sets of 9 and 6 frequencies in the M+B band,
is $<6 \times 10^{-4}$.  The probability of 4 K=4 type matches is
even smaller. Even if we restrict ourselves to the narrower
frequency interval of just the M band alone, the probability
of 5 matches is still smaller than 1.5 $\times 10^{-3}$.

We may therefore conclude, with more than 99\% statistical
confidence, that the two LC oscillate with the same 5 frequencies.
The similarity in the frequency values, within the
very small observational uncertainty limits, makes it very
likely that all the 5 had stable frequencies for at least the 16
days separating the two X-ray observations.

\subsection{The 0.75 mHz Feature}

The 0.75 mHz dominant feature in the PS of both LC deserves a
special comment. It is statistically significant in each
PS alone. It indicates clearly that an X-ray source emitting a
periodic or a quasi-periodic signal, contributed significantly
to the emission of the source at the two observation times.
In the PS of the Chandra data, this feature is unresolved
and the peak in the spectrum is found at the frequency 0.755 mHz.
This feature in Chandra PS is distinctly wider than other peaks,
indicating that it very likely an overlap of more than a single
frequency. Furthermore, our least squares fitting procedure
establishes that the fit to the data with two separate
frequencies, 0.728 and 0.782 mHz, is much better than with the
0.755 single frequency. The later is indeed the average of the two
separated  frequencies in the XMM-Newton LC. This claim is not
just the trivial statement that a fit to data of a function with
two free parameters is better than with one parameter only. The
fit to the Chandra data with the 2 frequencies is even better than
the fit with the 3 frequencies of the 3 most significant peaks in
the PS of Chandra.

The f2=0.728 mHz frequency is practically the same in the XMM and
the Chandra LC, although in XMM-Newton its amplitude is about
1/4th of what it was during the Chandra observations. The
corresponding period is P2=1371 s. The frequencies of the second
component of the 0.75 mHz feature in the two LC do not match each
other as well, but they are coincident with each other by the
criterion discussed in Section 6.1 with K=4 (see Table 1). The
cause of the mismatch here is probably in the 18000 s modulation
of the Chandra
LC, imposed by the decline of the emission at the end of that run
(Fig. 1d). The second frequency that we find in the 0.75 mHz
feature of
Chandra is close to 0.782 mHz, the frequency of the beat
of a 18000 s cycle with the 0.728 frequency. As the frequency of
the second component of the 0.75 mHz feature we take f3=0.763 mHz,
which is the value found in the least squares fitting process, as
well as in the PS of the XMM-Newton LC, where this component is
well resolved. The corresponding period
is P3=1310 s. Note that within a narrow margin of uncertainty in
the value of the frequencies, the following relation holds:

\begin{equation}
f1\simeq f3-f2
\end{equation}

The second harmonics of the 0.75 mHz feature is also
clearly present in the PS of the two LC. In the
XMM-Newton PS it is also resolved into 2 components of frequencies
f4=1.459 mHz and f5=1.504 mHz. The corresponding periods are
P4=685 sec and P5=665 sec. The first one satisfies the relation

\begin{equation}
f4 \simeq 2 \times f2
\end{equation}

 while the second one satisfies
\begin{equation}
 f5 \simeq f2+f3.
\end{equation}

The 3d and 4th harmonics of the 0.75 mHz feature are also
clearly present in the Chandra PS, each one as a single,
unresolved peak around the mean respective frequency.

\subsection{Energy dependence}
As described in Section 3, from measurements by the epic-pn camera
on board XMM-Newton we were able to extract 3 narrow band LC: LOW
(0.2-0.4 keV), MED (0.4-0.6 keV) and HIGH ($>$0.6 ksV). These
``color'' LC cover only 30000 second, rather than 36000 second
covered by the RGS camera. Their spectral resolution is therefore
somewhat lower than that of our average LC analyzed in the
previous sections. Figure 4a displays a blow-up of the PS of the
average LC shown in Figure 3a, around the frequency of the 0.75
mHz feature. Figures 4b, c, and d display the same frequency
interval in the power spectra of the LOW, MED and HIGH LC. The
feature is resolved into 2 components in the LOW PS where the 1310
component is higher than the 1371 sec one. The feature is
unresolved but appears as a single, broad peak in Figure 4c and
4d. Its profile is however clearly different in these two curves.
The peak frequency in the MED PS corresponds to P=1342 sec, the
peak in the HIGH PS is at the frequency corresponding to P=1367
sec. It therefore seems that there is a systematic change in the
ratio of the power in the two components. At low energy the 1310 s
cycle is dominating. At the mid-energy the two cycles contribute
equally to the power spectrum, while at our high energy band the
1371 sec cycle takes the lead.

With the data at hand we are unable to quantify the
statistical significance of this apparent trend. We do note,
however, that none of the other peaks in the PS of the 3 color LC
show similar variations, much less systematic ones, among the
different energy bands.

\begin{figure}
\includegraphics[width=87mm]{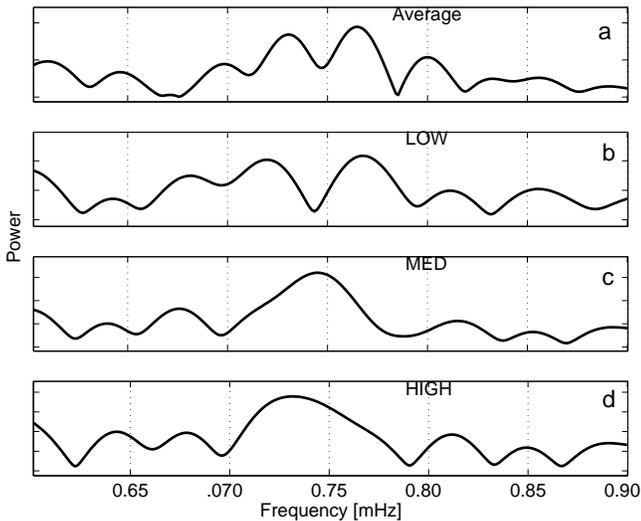}
\caption{a) A blow-up of the frequency region around the 0.75
mHz feature in the XMM-Newtin PS shown in Figure 1a.  b,c,d) The
same frequency region in the PS of the LOW, MED and HIGH XMM-Newton
LC.
}
\end{figure}

\section{Discussion}

The XMM-Newton and Chandra measurements reveal that
some 6 months after the outburst of nova V4743 Sgr, the
supersoft X-ray flux of this object oscillated with a set of
discrete frequencies. In fact, except for the decline in the
emission detected in March 2003, the temporal behavior of the
supersoft X-ray emission of the nova during the two observations
is well interpreted as consisting entirely of oscillations in 10-
20 discrete frequencies, superimposed on some DC
emission. Most of these frequencies are confined to the M
frequency band. All frequencies in the B band seem
to be higher harmonics of the more fundamental frequencies in the
M band.

\subsection{The 24000 sec cycle}

The period P1=24000 s associated with the most dominant
feature in the PS of the XMM-Newton LC is almost identical
to the optical photometric period discovered by Wagner et
al. (2003). It is tempting to identify the X-ray periodicity
with the optical one but some caution is still called for since
less than two complete cycles of this periodicity have been
recorded in the X-ray so far. Wagner et al suggested that
this is the binary period of the system. The question of the
possible identification of the 24000 s with the orbital period
of the nova system is of course very important. Comparison
of the 2 X-ray LC indicate that if the 24000 s is indeed
the binary period, then the decline observed with Chandra
in March of 2003 (Ness et al. 2003, Starrfield et al. 2003)
is not an eclipse related to the binary revolution. If this
were true, the system would have undergone considerable
geometrical changes
between March 20 and April 4, which seems highly unlikely.

\subsection{The two major periodicities}

The P2 and P3 periodicities that constitute the 0.75 mHz 
feature in the PS of both LC, may be the signatures of non 
radial pulsations of the white dwarf of this nova system (see 
next section). However, their prominence in the two LC and 
relation (1) of Section 6.3 may indicate that their origin is 
different from that of the other periodicities in the LCs. 
We suggest that one of these 
two periods is the spin period of the WD. A cycle 
of $\simeq$1300 s is indeed a typical value of spin periods 
in IP systems (Hellier 2001). In Section 6.4 we showed that the ratio
between the power of the P2 and P3 periodicities is energy dependent.
This lends some support to the suggestion that these two frequencies
originate in two different mechanisms. If one of them is a member in the
family of the other major oscillations of the system, as suggested in
the next section, the other is foreign to it, and the spin of the WD is
a natural candidate to be its source.

If the 24000 s is the orbital period of the system, from
relation (1) in Section 6.3 we see that the second period
of the 0.75 mHz feature could be the orbital sideband cycle of the WD.
For a prograde spin,
the sidereal spin period must be the shorter of the two, namely
P3=1310 s. Signals of the spin period and of its sideband
have been detected in the light curves of a number of Intermediate Polar
CVs (Warner 1995). An example is FO Aqr
whose X-ray light curve observed with Ginga varied with
two periods (Hellier 1993) . In fact even the numerical value
of that WD spin period, 1254 s, and of its orbital period,
17460 s, are rather similar to our case.

Two major possibilities, well explained in
 Hellier (2001), may explain the orbital sideband
oscillations are: a) diskless accretion by a magnetic WD,
and b) reprocessing of beamed radiation from the spinning
WD by an element in the binary system that is fixed in the
orbital rotating frame. Even with the rich spectral information
 we have obtained in this observation (Ness et al. 2003),
 we do not find indications to discriminate between these two possibilities
 by analysing the spectra and variations in the depth of lines.

We shall not attempt to provide a model for
the temporal characteristics of the X-ray emission of V4743
Sgr described in this work. We note, however, that on the basis of the
sideband
models for the f2 and f3 frequencies it is difficult to explain
the presence of f5 in the PS of XMM-Newton, which satisfies
relation (3) of Section 6: f5=f2+f3.

\subsection{Other cyclic variations}

Two other periodicities, P6 and P7 (see Table 1), persisted in the
LC of the star for 16 days or more (Section 4).
Taking for each the average value, weighted by the respective resolving
powers, of the two measurements  by Chandra and by
XMM-Newton, we adopt P6=3218 sec and P7=5602 sec.

The data do not allow establishing at a
statistically significant level the presence of additional
periodic oscillations. It is however likely that some of the other
frequencies found the two LC, e.g. the one corresponding to the
2nd highest peak in the XMM-Newton LC at P=9860 second, are also
signals of periodic or quasi-periodic oscillations of the X-ray
emission of the star.

Several different and simultaneous periodic or quasi periodic
oscillations in the X-ray light curve have never been
observed before in classical nova systems. It is tempting to
suggest that at least some of the observed X-ray variability is
due to white dwarf pulsations. The signature of white
dwarf pulsations in the X-ray emission of classical novae has
already been found for V1494 Aql (Drake et
al 2003). It will be important to assess whether some of
the periodicities we have detected are indeed due to nonradial
pulsations of the hot WD, whose atmosphere must be
extremely rich in oxygen, since the CNO cycle is on going in a
thin shell underneath the atmosphere. Starrfield et al (1985) show
that instabilities may occur due to ionization of carbon and
oxygen and may give rise to pulsations of time scales of the order
of the periodicities that we have discovered in the M band
frequency range. These calculations, performed for hot stars
evolving from the asymptotic giant branch to the white dwarf
cooling sequence, may not be directly relevant to our case since
the effective temperature of
V4743 Sgr in these observations was above  500,000 K (see
results obtained by Petz et al. 2005, and by Rauch \& Orio
2004, Orio et al. 2005 with two different methods and models).
They do point, however, toward the possibility that a
similar mechanism operating on another atomic species at
a higher excitation state, may produce the observed cyclic
variations of our source.

Non radial pulsations of a similar time scale were also
detected in the light of the hot, very evolved WD PG 1159035 and
RXJ 2117.1+3412, with periods around 230-830 s
(McGraw et al. 1979, Vauclair et al. 1993). WD that are nuclei of
planetary nebulae have also been observed to undergo
nonradial pulsations (Grauer \& Bond 1984).
If confirmed, nonradial pulsations of novae WD offer a
new way to understand the chemical composition and temperatures of
the WD and will provide new means to study
the system evolution as well. X-ray observations of post-
outburst novae may therefore be even more rewarding than
previously thought.

\section{Summary and Conclusions}

Variability by up to 40\% of the total X-ray flux, has been discovered
in the X-ray light curve of the classical nova V4743 Sgr 6 and 6.5
months after the optical maximum. The light curve was sampled in
March 2003 for 8 hours by Chandra,
and 16 days later for 10 hours with XMM-Newton. It was
found to vary intensely on all time scales between the duration of
the observations and about 300 s. One of these
variations seems to correspond to a period discovered at
optical wavelengths, which was identified as the orbital period of
the system.

We have found at least 5 periodicities in addition to
the suspected orbital one. They appear to be the signals of
oscillations with frequencies 
that persisted in the LC for more than
16 days. Other periodicities or quasi-periods characterize the
LC as well. Some of the periods are likely to be reflections
of nonradial pulsations of the atmosphere of the hot WD of
the nova. Two of the periodic variations with the periods
1371 s and 1310 s are particularly intense. One of them
may be the spin period of the WD of this nova system. The
frequency that is the sum of these two frequencies has
also been identified in the LC.

The results of our time series analysis of the
XMM-Newton and Chandra observations underline the value of
further, more intense time resolved photometry of this interesting
nova and of classical novae in general, in the X-ray,
as well as in the optical region of the spectrum.

\section*{Acknowledgments}

We are very grateful to the  XMM-Newton Project Scientist Fred Jansen for
scheduling the observations, to Pedro Rodriguez of the  XMM-Newton ESA team
at Vilspa for  help, and to Steve Snowden of the NASA-GOF for advice.
CV studies at the Wise Observatory are supported by the Israel Science
Foundation. M. Orio acknowledges financial support of the University
of Wisconsin Graduate School and the NASA funding for the XMM-Newton program.
S. Starrfield and J.U. Ness acknowledge support from NSF and
CHANDRA funding at ASU. We also thank the referee, Dean Townsley, for
some very useful suggestions.

\label{lastpage}

\end{document}